\documentclass[fleqn,10pt]{wlscirep}

\usepackage{xr-hyper}

\usepackage[utf8]{inputenc}
\usepackage[T1]{fontenc}
\usepackage{lineno}
\usepackage{siunitx}
\usepackage{xcolor}
\usepackage{listings}
\usepackage{amsmath}
\usepackage{xr}
\usepackage{hyperref}
	
\hypersetup{pdfcreator={Me}}

\title{Classifying soft self-assembled materials via unsupervised machine learning of defects}

\author[1]{Andrea Gardin}
\author[2]{Claudio Perego}
\author[2]{Giovanni Doni}
\author[1,2,*]{Giovanni M. Pavan}
\affil[1]{Department of Applied Science and Technology, Politecnico di Torino, Corso Duca degli Abruzzi 24, 10129 Torino, Italy}
\affil[2]{Department of Innovative Technologies, University of Applied Sciences and Arts of Southern Switzerland, Polo Universitario Lugano, Campus Est, Via la Santa 1, 6962 Lugano-Viganello, Switzerland}

\affil[*]{Corresponding author: Giovanni M. Pavan (giovanni.pavan@polito.it)}

\usepackage{xr-hyper}

\makeatletter
\newcommand*{\addFileDependency}[1]{% argument=file name and extension
  \typeout{(#1)}
  \@addtofilelist{#1}
  \IfFileExists{#1}{}{\typeout{No file #1.}}
}
\makeatother

\newcommand*{\myexternaldocument}[1]{%
    \externaldocument{#1}%
    \addFileDependency{#1.tex}%
    \addFileDependency{#1.aux}%
}

\myexternaldocument{SI_comparison}

\usepackage[version=3]{mhchem} % Formula subscripts using \ce{}
\usepackage{amssymb}
\usepackage{xcolor}
\usepackage{nameref}
\usepackage{amsmath}

\begin{abstract}
Unlike molecular crystals, soft self-assembled fibres, micelles, vesicles, etc., exhibit a certain order in the arrangement of their constitutive monomers, but also high structural dynamicity and variability. Defects and disordered local domains that continuously form-and-repair in their structures impart to such materials unique adaptive and dynamical properties, which make them, \textit{e.g.}, capable to communicate with each other. However, objective criteria to compare such complex dynamical features and to classify soft supramolecular materials are non-trivial to attain. Here we show a data-driven workflow allowing us to achieve this goal. Building on unsupervised clustering of Smooth Overlap of Atomic Position (SOAP) data obtained from equilibrium molecular dynamics simulations, we can compare a variety of soft supramolecular assemblies \textit{via} a robust SOAP metric. This provides us with a data-driven “defectometer” to classify different types of supramolecular materials based on the structural dynamics of the ordered/disordered local molecular environments that statistically emerge within them.

\end{abstract}

\begin{document}
\flushbottom
\maketitle

\section{Introduction}

Supramolecular structures, composed of molecular units that self-assemble \textit{via} non-covalent interactions, represent the key substrate for biological systems (membranes, micelles, protein fibres, etc.), and for new types of self-healing, stimuli-responsive and bioinspired materials.\cite{Aida2012,Brunsveld2001,degreef2008,boekhoven2014}  %%% REFS su supramolecular stuff
Among many interesting features, the highest potential of soft supramolecular materials lies in their intrinsically dynamic character.\cite{lehn1993} The self-assembled monomers continuously exchange within and in-and-out these materials,\cite{bochicchio2017natcommBTA,gasparotto2019identifying} controlling how they communicate with each other at the equilibrium,\cite{demarco2021controlling,crippa2021} how they respond to external stimuli,\cite{torchi2018dynamics} and how they behave out-of-equilibrium.\cite{bochicchio2019defects} 
Such dynamical features resulted in a set of intriguing properties that, on the one hand, mimic natural ones, crucial for the functioning of biological tissues (\textit{e.g.}, self-healing, chemotacticity, molecular transport, etc.)\cite{Li2008,Matson2011,Bastings2013,Nagel2017,lionello2021} and, on the other hand, are promising features for the design of new functional materials and nano-technologies\cite{Yan2012,Davis2002,Lehn2005,Aida2012,lancia2019,babu2021,xiu2021,gentile2021} % + example on liquid crystals ? and/or micelles? or other stupid aggregates? 

\begin{figure}[h!]
    \centering
    \includegraphics[width=0.9\textwidth]{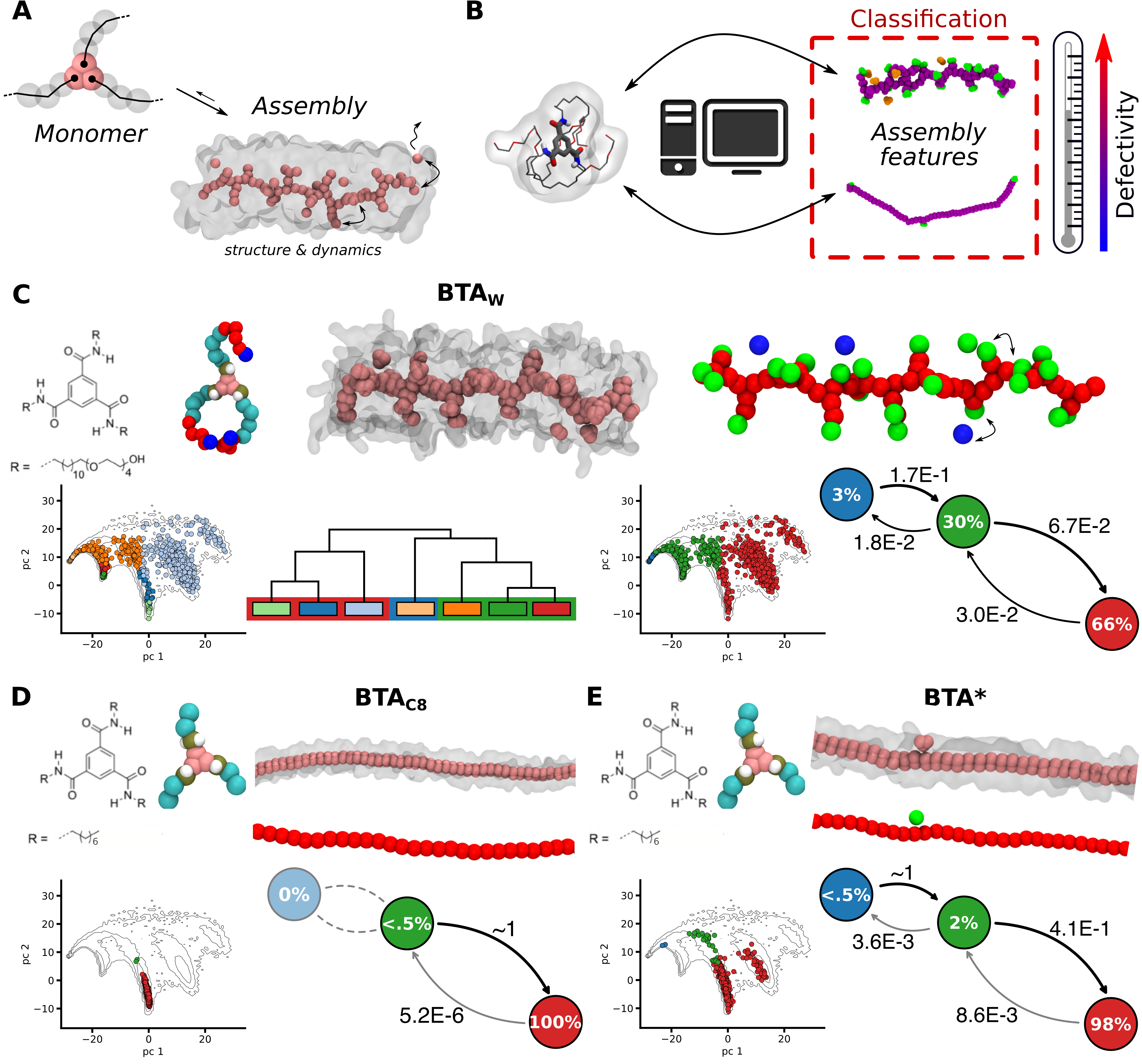}
    \caption{Classification of supramolecular polymers based on the structural/dynamical features of their molecular motifs. (A) Scheme showing an example of monomer self-assembling into supramolecular 1D fibres having a characteristic level of order/disorder which determines its structural/dynamical features (\textit{e.g.}, BTA).\cite{bochicchio2017natcommBTA,gasparotto2019identifying,bochicchio2017acsn} (B) A key question is how to design the monomers to obtain assemblies with controlled features. While ML techniques can be useful to this end, a first necessary step is developing a classification approach to compare different supramolecular structures in an unbiased way.
    (C) Analysis of the water-soluble BTA$_W$ fibre. Top-left: chemical structure and CG model of BTA$_W$ monomers and equilibrium CG-MD snapshot of a BTA$_W$ fibre.\cite{bochicchio2017natcommBTA,gasparotto2019identifying,bochicchio2017acsn} Bottom-left: PCA scatter-plot of molecular motifs identified \textit{via} SOAP (projected on the first two PCs: PC1 and PC2). %A representative subset of SOAP vectors specific to the BTA$_W$ system is shown as coloured circles 
    Each colour corresponds to a different motif detected in the BTA$_W$ fibre (the underlying contour plot shows the distribution of all SOAP vectors sampled in the three compared BTA fibres. The dendrogram indicates how the various detected motifs (microclusters: smaller rectangles) relate to each other, and how these can be merged into higher-level motifs (macroclusters: red, blue and green rectangles). Top-right: colouring of monomer centres in an example equilibrium CG-MD snapshot based on the macro-state. Bottom-right: PCA scatter-plot projection for BTA$_W$ coloured based on the macrocluster. Right: dynamic interconversion diagram, reporting the relative probability of transition between various states, and the relative state population.
    % Top-right: matrix of transition probability between (higher level) MMs, snapshot of the fibre model (displaying only the monomer centers colored according to their MM), and interconversion diagram, reporting the probability of transition between states.
    % Bottom-right: scatter-plot of (higher-level) MMs, and histogram of MM population. 
    (D,E) Analogous SOAP+PAMM characterization for the BTA$_{C8}$ (D) and BTA$^*$ (E) fibres based on the macroclusters. %Asterisks indicate rare fluctuation states.
    % \cla{Cosa sono i fluctuant states? Bisogna specificarlo da qualche parte}.
    % reference monomer chemical formula and CG-beads representation, a snapshot of the assembled infinite BTA-variant fibre with the corresponding structural motifs highlighted as the colors of the cluster analysis, a scatter plot of the specific fibre dataset (showing only the two first principal components) on top of a density isoline plot of the whole dataset combined, and finally the interconversion probability matrix for the clusters and the clusters histograms for every BTA fibre model considered. d) Schematic visualization of the transitions computed from the interconversion probability matrices.
    } 
    \label{fig:fig1}
\end{figure}

A major goal in the study of self-assembled architectures is understanding how changes in the structure of the self-assembling building blocks (input) affect the overall properties of the supramolecular assembled structure (output) (Fig. \ref{fig:fig1}). Gaining such knowledge would pave the way toward the rational design of new functional materials with controlled properties, reducing the costs of trial-and-error synthesis.
However, because of the complexity and the dynamical, multiform nature of such systems,\cite{Capito2008,Marchetti2013,Chen2011,Sanchez2012} %and the fact that the monomers interact not only with each other, but also with the solvent in which they are immersed, 
a direct connection between the properties of supramolecular materials with the features of the constituent monomers remains often impossible to attain.

Molecular simulations play a paramount role in this framework, providing high-dimensional data on the structure and dynamics of supramolecular assemblies, with detailed chemical-physical information on the factors that control the dynamical features of these materials.\cite{garzoni2016,tantakitti2016,bochicchio2017acsn,bochicchio2017natcommBTA,bochicchio2019defects,gasparotto2019identifying,carter-fenk2019,souza2021}

However, analysing and interpreting the high-dimensional, high-detail data produced by molecular simulations can be challenging, particularly in supramolecular systems. %, data is that of finding reliable and rigorous descriptors, also named collective variables (CV), that isolate the relevant degrees of freedom of a complex, highly dimensional system,  and characterize the structural arrangement of the monomers in the supramolecular architecture. 
In such complex dynamical structures, irregularities and defects play a crucial role,\cite{bochicchio2017natcommBTA,torchi2018dynamics,bochicchio2019defects,demarco2021controlling,lionello2021} demanding for system descriptors capable to translate the simulation data into a reliable classification of the molecular structure and dynamics charactristic of these systems. %Thus, to provide a comprehensive understanding of these systems, CVs must be able to characterize and classify structural defects as well. 

"Human-based" analyses rely on low-dimensional descriptors, defined based on the experience and on those features directly readable by visual inspection. This could lead to biased predictions, affected by the choice of such low-dimensional descriptors, possibly overlooking important degrees of freedom of the system. To overcome these limitations, one can employ data-driven, Machine-Learning (ML) approaches such as unsupervised dimensional reduction, and pattern recognition.\cite{ceriotti2011simplifying,artrith2017efficient,cheng2020mapping,deringer2021} These techniques fully exploit the rich, high-dimensional information provided by simulations, extracting crucial patterns and correlations, that enable a more effective system characterization.\cite{ceriotti2019,wang2019,cheng2020mapping}%non so bene cosa citare
In particular, data-driven analyses allow to effectively monitor the environment of each atom/molecule in a system, providing ``fingerprints" that classify the mutual arrangement of atomic/molecular entities.\cite{behler2011,Rupp2012,Faber2015,huo2018arxiv,bartok2013} The Smooth Overlap of Atomic Position (SOAP)\cite{bartok2013}, proved to be very efficient in encoding the information of molecular environments into a rich, high-dimensional and agnostic (i.e.~independent from \textit{a priori} knowledge of the system) description.\cite{de2016comparing,gasparotto2018PAMM,engel2018mapping} %Moreover, by inspecting the mathematical basis and requirements of these descriptors it was possible to formulate a unifying theory of these atom-density representations.\cite{willat2019braket}
Coupling SOAP analysis with unsupervised clustering enables a robust characterization of the ordered/disordered arrangement in molecular systems, as demonstrated for Lennard-Jones clusters, peptides\cite{gasparotto2018PAMM} and water.\cite{gasparotto2014a,monserrat2020liquid,capelli2021arx} Recently, combining SOAP descriptors and unsupervised clustering, we reconstructed the complex structure and dynamics of molecules in supramolecular materials, tracking \textit{e.g.} the defect formation and evolution.\cite{gasparotto2019identifying} This data-driven approach enables a rigorous and unbiased classification of soft, supramolecular structures.\cite{capelli2021,bian2021}

In the present work we push the data-driven comparability between supramolecular assemblies to the limit. Building on SOAP calculation and unsupervised clustering, as well as on a SOAP-based distance definition, we construct a high-dimensional SOAP metric, which can be used as a "defectometer", measuring and comparing different structures based on their local molecular environments. This allows us to compare different types of assemblies, such as supramolecular fibres, micelles, layers, nanoparticles, etc. We obtain a general, unbiased classification of supramolecular systems that are profoundly different among each other, capturing structural and dynamic aspects that are hardly readable with standard descriptors.

\section{Results and discussion}\label{sec:res}

\subsection{Comparing variants of a supramolecular polymer}

Recently, it has been shown that unsupervised clustering of SOAP data from equilibrium MD of 1,3,5-Benzenetricarboxamides (BTA) supramolecular polymers allows reconstructing the structural dynamics of such complex assemblies.\cite{gasparotto2019identifying} 
In particular, defining a SOAP vector in the center of each BTA monomer allows to retrieve detailed information on the level of order/disorder (local and average), %order dis-homogeneity, 
as well as on the internal dynamics of monomers in a supramolecular polymer.
Considering different variants of BTA supramolecular fibres, and combining the SOAP data obtained for each system, generates a unique data-set gathering all possible configurations (each one identified by a SOAP vector) visited by the monomers in the different BTA fibres. This guarantees that the SOAP vectors associated to each monomers in the different assemblies belong to the same high-dimensional feature space, allowing to compare the fibre variants.% by projecting their SOAP data on the global SOAP feature space.
\cite{gasparotto2019identifying} 
The resolution of the molecular models is relevant in this sense, as it implicitly determines the accuracy by which, \textit{e.g.}, the monomer-monomer interactions, the monomers' flexibility, etc., are described.
Coarse-grained (CG) models with a resolution <5 \AA --- developed based on the Martini force-field,\cite{marrink2007martini} and optimized to match the behaviour of the respective all-atom (AA) models --- were proven accurate enough to capture all essential information while guaranteeing sufficient sampling of the assembled fibres in equilibrium.\cite{bochicchio2017acsn,bochicchio2017natcommBTA,gasparotto2019identifying} 

With such CG models, it was demonstrated that water-soluble BTA polymers, composed of BTA$_{W}$ monomers with amphiphilic arms, possess a variegated structure with a rich and diverse set of monomeric states, in continuous exchange and dynamic interconnection with each other. The SOAP data of the BTA$_{W}$ fibre were analysed \textit{via} the Probabilistic Analysis of Molecular Motifs (PAMM) unsupervised clustering (see Methods section for details).\cite{gasparotto2019identifying,gasparotto2018PAMM} Shown in Fig. \ref{fig:fig1}C, PAMM identifies different molecular motifs (microstates) in the SOAP data, shown by different colours in the PCA plot. The hierarchical structural/dynamical interconnections between microstates are captured by the related dendrogram (left). As the SOAP data are collected at different snapshots of equilibrium MD trajectories, colours that are found close to each other in the dendrogram may identify states that are similar from a structural point of view and/or quickly exchanging monomers with each other. The dendrogram also allows to perform a systematic CG of the classification, identifying the dominant motifs (macrostates) in the BTA$_{W}$ fibre (Fig. \ref{fig:fig1}C, right): the ordered/persistent interior (bulk) of the fibre (in red: $\sim66 \%$ of the monomers), the stacking defects ($\sim30 \%$ of the monomers, in green: bound to the stack only by one side), and the monomers which are adsorbed on the fibre surface, moving from one defect to another one (in blue: population $\sim3 \%$ of the monomers). From the frequency at which the monomers change state during the equilibrium MD, we can estimate the relative transition probabilities between the various states. The BTA$_{W}$ fibre turns out to possess a very rich and diverse internal dynamics.
In comparison, the BTA$_{C8}$ variant, with shorter carbon side-chains, produce straight and substantially defect-free fibres (Fig. \ref{fig:fig1}D). Interestingly, artificially reducing the directional interactions between the monomers a new BTA$^{*}$ fibre variant is defined, where a fraction ($\sim2 \%$) of monomers form defects that are continuously created-and-repaired.\cite{gasparotto2019identifying} 

It is worth noting that such a direct comparison between fibre variants is possible since all SOAP vectors sampled in the compared systems belong to the same high-dimensional feature space ($324$-dimensional, see Methods for details). This approach allows to identify defects (\textit{i.e.}, disordered domains) in the supramolecular fibres, to track their emergence/disappearance, %repairment,
and to use them to compare fibre variants. Noteworthy, as the concept of defects is elusive in such statistical/dynamical structures, this approach is general and flexible, as the definition of the molecular motifs that statistically populate the assemblies, and thus of defects, is exquisitely data-driven. This analysis can be thus extended to other families of assemblies. As a first proof-of-concept, we proceed by comparing variants of different supramolecular polymer families.

\begin{figure}[h!]
    \centering
    \includegraphics[width=0.9\textwidth]{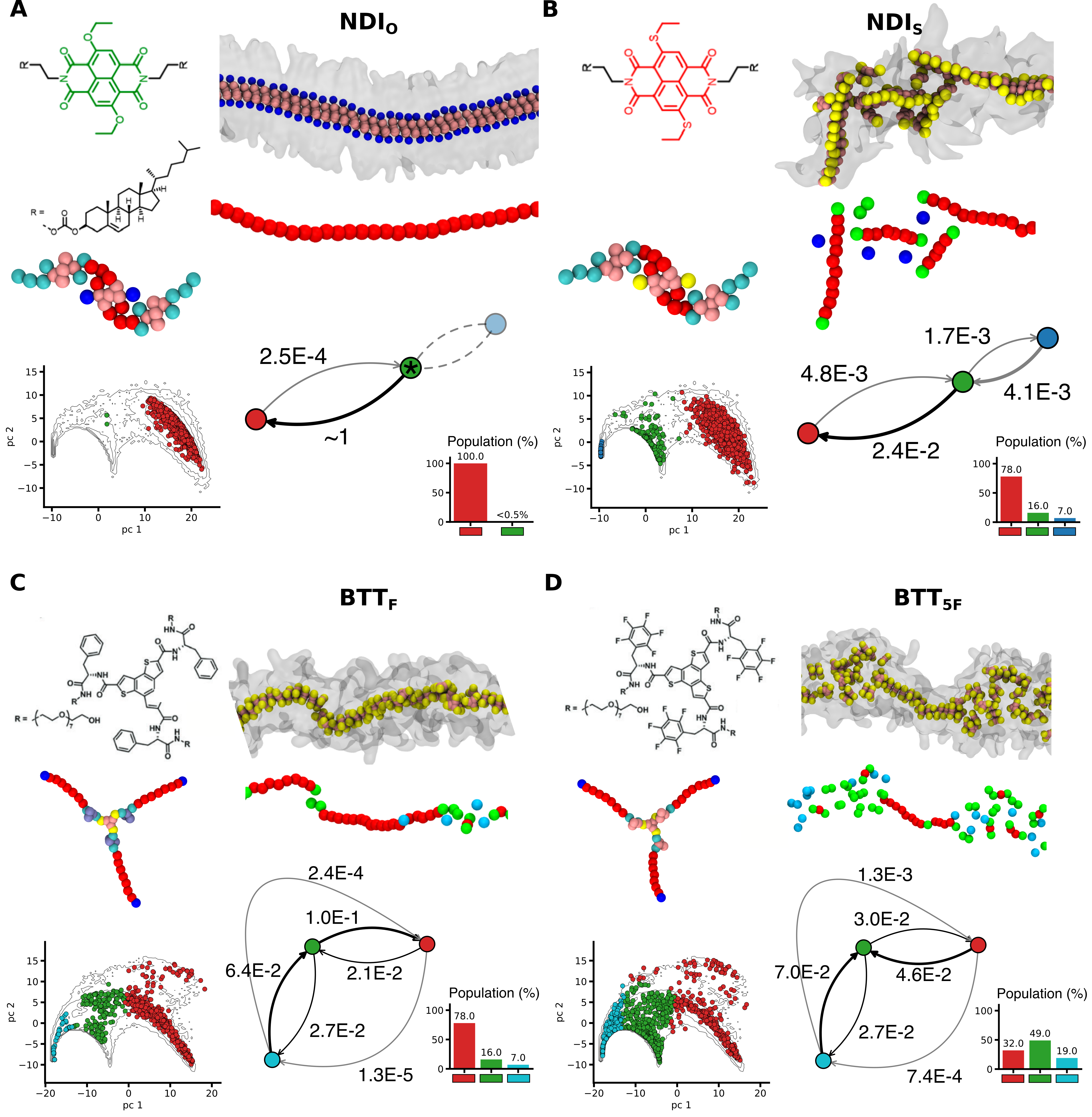}
    \caption{Comparing variants of supramolecular polymers. (A-D) SOAP and PAMM analysis on NDI and BTT supramolecular polymer variants. (A) NDI$_O$ monomer, (B) NDI$_S$ monomer, (C) BTT$_F$ monomer, and (D) BTT$_{5F}$ monomer. Each panel summarizes the results of the analysis for a single system variant, showing (from top-left): chemical structure and CG model of the monomer; equilibrium CG-MD snapshot of the assembly, and view of the same with monomer centers coloured according to the identified molecular motifs; scatter-plot showing the PCA (projected on the first two PCs) of SOAP descriptors, coloured according to the motif. Interconversion diagram and population histogram. Asterisks indicate rare fluctuation states, with residence time $\sim0$ (\textit{e.g.}, green cluster in NDI$_O$).
    } 
    \label{fig:fig2}
\end{figure}

We focus on two other families of supramolecular polymers, formed by monomers based on naphthalene diimide (NDI)\cite{sarkar2020NDImonomer} (Fig. \ref{fig:fig2}A,B) and benzotrithiophen cores (BTT)\cite{casellas2018btt} (Fig. \ref{fig:fig2}C,D). We compare two variants per-family, NDI$_O$ \textit{vs.} NDI$_S$, and BTT$_F$ \textit{vs.} BTT$_{5F}$, for which we employ reliable CG models having analogous resolution of the BTA models used in the analyses of Fig. \ref{fig:fig1}.\cite{sarkar2020NDImonomer,casellas2018btt} We repeated the analysis performed for BTAs to compare the NDI$_O$ and NDI$_S$ fibre variants, building a unique SOAP dataset from the equilibrium MD of the two systems (Fig. \ref{fig:fig2}A-B). Also in this case, we define one SOAP vector in the center of each monomer, computed over MD trajectories having the same sampling frequency and length % same-length equilibrium MD simulations, sampling the same number of MD snapshots with the same time-lapse frequency 
(see Methods section for complete details). We then performed PCA of the SOAP dataset, and used PAMM to identify the molecular states. Also in this case, PAMM finds three dominant monomeric states in the NDI fibres (Fig. \ref{fig:fig2}A-B: in red, green and blue). 
Indeed, this analysis of NDI variants evidences a picture similar to that of BTAs. Already the 2D PCA projections of the SOAP vectors highlight different structural motifs in the two fibres (the separated spots in the scatter-plots of Fig.~\ref{fig:fig2}A \textit{vs.} B). One variant, NDI$_{0}$, exhibits a well-ordered structure with all monomers belonging to the fibre backbone (Fig. \ref{fig:fig2}A: in red), apart from rare fluctuations. The slight change in the NDI$_{S}$ monomer structure (the replacement of two oxygen with two sulfur atoms) produces instead a more disordered fibre, rich of defects (Fig. \ref{fig:fig2}B, in green: $\sim16 \%$), and of adsorbed monomers which travel between defects within the assembly (in blue: $\sim7 \%$). 
The same protocol was used for the BTT fibres (Fig. \ref{fig:fig2}C-D), showing again three molecular states, and providing a picture similar picture to that of BTA and NDI systems.  %Both in the NDI and in the BTT cases we selected a single-center SOAP definition, centered on the COG of the monomers core residue (Fig. \ref{fig:fig2}).
In particular, the BTT$_{F}$ molecule tends to form a much ordered fibre than the BTT$_{5F}$ variant, in agreement with what is known experimentally for these systems.\cite{casellas2018btt} A difference that emerges from these analyses, is that BTT fibres possess a more diverse structure than NDI fibres, which appear as more ordered in comparison. In particular, BTT$_{5F}$ exhibits the highest occurrence of disordered monomeric domains ($\sim49 \%$ of the monomers are in defected state and $\sim19 \%$ are adsorbed/travelling monomers). The ordered monomers in the BTT$_{5F}$ fibres are only $\sim32\%$ ($\sim78\%$ in BTT$_{F}$), and considerable dynamic interconversions are observed between all monomeric states (Fig. \ref{fig:fig2}C-D). %This can be inferred from the larger number of dynamic interconnections between the states of Fig. \ref{fig:fig2}C-D, and by the fact that, \textit{e.g.}, the BTT$_{5F}$ fibres start to be "more disordered than ordered", with only $\sim32 \%$ of the monomers that populate ordered states \textit{vs.} $\sim49 \%$ of the monomers which are found in defected states, and $\sim19 \%$ of reshuffling monomers. 

With this analysis, in most of the studied systems we qualitatively detect the statistical presence (or formation) of defects. This is an important feature since, as demonstrated for BTA supramolecular polymers,\cite{bochicchio2017natcommBTA,demarco2021controlling ,gasparotto2019identifying} the possibility to create and repair defects in the assembly is crucial for their dynamic properties.\cite{torchi2018dynamics}
The analyses of Figs. \ref{fig:fig1} and \ref{fig:fig2} demonstrate that this approach is versatile, and it can be used to compare fibre variants within the same supramolecular polymer family. However, at this stage, the analysis does not allow us to unambiguously compare between the different families. 
In particular, since the definition of defects emerge from the data contained in separate SOAP datasets, it is not clear to what extent the defects in the BTA fibres are comparable to the BTT or NDI ones. The SOAP dataset used to compare the BTA variants in Fig. \ref{fig:fig1}, in fact, does not contain information the monomer states in the NDI and BTT fibres. To compare these supramolecular fibres in a more objective and quantitative way, a further step is required. % The comparison between these supramolecular fibres remains thus arbitrary and qualitative at this level. This requires one more step. 

\begin{figure}[h!]
    \centering
    \includegraphics[width=0.9\textwidth]{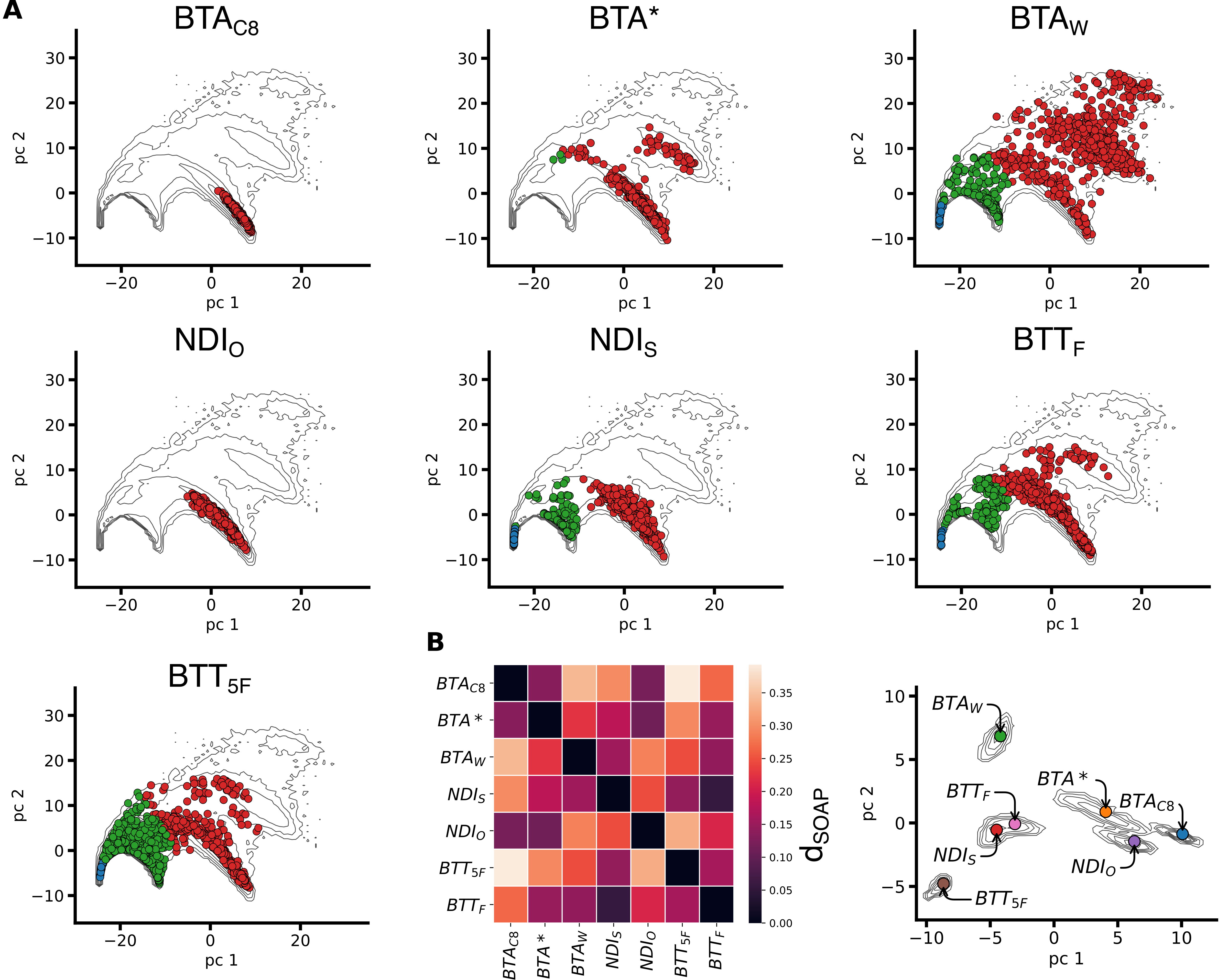}
    \caption{Comparing 1D assemblies. (A) Scatter-plots of molecular motifs (macroclusters) detected for the different analysed supramolecular polymers. The PCA projected on the first two PCs (PC1 and PC2) of the of SOAP vectors are shown, coloured based on the identified (PAMM) macroclusters. The scatter plots are embedded in a global contour plot of the complete dataset (SOAP vectors of all the seven 1D supramolecular fibres). (B) Distance ($d_{\mathrm{SOAP}}$) matrix (left) built from the \textit{simulation}-averages of the SOAP vectors (Eq. \ref{eq:kerdistance} in the Methods section). The colour scale indicates the distance in the high-dimensional SOAP feature space between the various assemblies. Right) Contour plot of the \textit{frame}-average distributions computed from the PCs of the SOAP vectors. The coloured dots represent the \textit{simulation}-averages for each 1D fibre.
    % Structural motifs comparison of the supramolecular fibres considered in our work. a) Scatter plots of the two first principal components of the SOAP analysis for each fibre family members on top of a density isoline plot of the whole combined dataset, colors encode the overall family structural motifs computed from a PAMM cluster analysis. b) SOAP average densities isolines plot projected onto theis two first principal components. The scatter plot on top represents the mean values of the SOAP averages for each fibre aggregates. Distance matrix computed from the mean of the SOAP average kernel induced pseudometric showing relationships occurring between the overall structural representation of each fibre aggregates.
    } 
    \label{fig:fig3}
\end{figure}

\subsection{Comparing different types of 1D supramolecular polymers}
As described above, by processing a comprehensive dataset containing SOAP vectors sampled via MD simulation of different systems, one can build a framework to rigorously and unambiguously compare different supramolecular systems, based on the identified molecular motifs. The idea is to retain in the SOAP analysis those relevant features that are common to all the systems (a sort of "common molecular denominator").\cite{gasparotto2019identifying} In this case, we consider only the monomer centers in the SOAP calculation (see the Methods section), with two advantages: (i) this has been shown to retain rich information on the monomer arrangement in the fibres, and thus on their supramolecular structure and dynamics.\cite{demarco2021controlling,gasparotto2019identifying} (ii) This makes the analysis very general/abstract --- all assemblies are composed of mutually interacting monomers ---, opening the possibility to compare, not only variants of a supramolecular fibre, but also widely different assemblies.

To this end, we thus built a global dataset containing all the SOAP vectors sampled \textit{via} the equilibrium MD of the systems simulated for the analyses of previous section (3 BTA, 2 NDI and 2 BTT fibres), obtaining the global contour plot reported in Fig.~\ref{fig:fig3}A. 
We repeated the PCA and PAMM analyses over this dataset, identifying the main molecular motifs shared by all these one-dimensional assemblies (Fig.~\ref{fig:fig3}A).
In Fig. \ref{fig:fig3}A, we display the PCA scatter-plots (first two PCs) for all 1D assemblies. As in the analysis of Figs.~\ref{fig:fig1} and \ref{fig:fig2}, the three main structural clusters identified by PAMM --- backbone (in red), defects (in green) and adsorbed/surface-diffusing monomers (in blue) --- are preserved. 
In Fig.~\ref{fig:fig3}A, the PCA scatter-plots indicate the SOAP distributions (states) sampled by the individual systems in the global SOAP dataset. 
These scatter-plots can represent characteristic ``fingerprints'' of the supramolecular structures, indicating which  molecular states are populated in each system. 
Qualitatively, such fingerprints provide an information of similarity between systems, in terms of structural arrangement and dynamicity of the monomers. % and structural dynamicity. 
However, since PCA scatter-plots are the result of dimensionality reduction, they can provide a distorted picture, where dimensions potentially relevant in the overall feature space may be neglected. %, and/or data lying on different dimensions may result superimposed. 

To reach a more quantitative insight, we employed a SOAP-based metric that allows comparing complex supramolecular systems based on the average SOAP spectra of the monomers forming the assemblies (\textit{i.e.}, based on the structural/dynamical features of the local environment that, on average, surrounds the monomers). Proven useful to compare complex interacting molecular systems, such as, \textit{e.g.}, lipid bilayers,\cite{capelli2021} or liquid aqueous systems,\cite{capelli2021arx} we build on such data-driven metric to quantify the similarity between the SOAP data associated to each fibre (see Methods for details).%, and thus between the various supramolecular polymers 
For each frame sampled by an MD simulation, the SOAP vectors associated to the ``local" environments surrounding all monomers are averaged into a single SOAP spectrum (the \textit{frame}-average, Eq. \ref{eq:psframeave}). These frame spectra are then averaged along the MD trajectory, obtaining an average SOAP spectrum, which is characteristic of a given assembly and of the molecular motifs that populate it at the equilibrium (the \textit{simulation}-average, Eq. \ref{eq:pssimave}). 
We can thus assess the similarity among the different 1D assemblies in a more rigorous way, by employing a SOAP-induced metric\cite{de2016comparing,capelli2021,capelli2021arx} (Eq. \ref{eq:kerdistance}) to compute the distance ($d_{\mathrm{SOAP}}$) between the SOAP \textit{simulation}-averages of the various systems. The result of this analysis is the $d_{\mathrm{SOAP}}$ distance matrix in Fig.\ref{fig:fig3}B (left). The \textit{frame}- and \textit{simulation}-averages associated to each system are projected onto the first two PCs in the plot of Fig.\ref{fig:fig3}B (right).%, as it gives a bounded measure of similarities among them.
The off-diagonal values in the $d_{\mathrm{SOAP}}$ matrix (Fig.\ref{fig:fig3}B, left), allow classifying the assemblies under investigation in the global SOAP space. The darker is the colour of the entry, the lower is the $d_{\mathrm{SOAP}}$ between the assemblies, indicating their similarity (in terms of molecular environments). The $d_{\mathrm{SOAP}}$ matrix confirms the qualitative indications provided by the scatter-plots of Fig. \ref{fig:fig3}A: the most ordered supramolecular polymers, namely, BTA$_{C8}$, BTA$^*$, and NDI$_O$, are similar and mainly populated by ordered, backbone-like molecular domains (Fig. \ref{fig:fig3}A: in red). In these systems, the defect states are just sparsely populated, and located very close to the ordered domains --- this is consistent with the fact that defects are relatively rare fluctuations %just rarely generated as statistical states
, which are readily repaired (consistently with, \textit{e.g.}, Fig.~\ref{fig:fig1}E).
The remaining 1D systems, containing a higher defect concentration (Figs. \ref{fig:fig1} and \ref{fig:fig2}), are more or less distant from the previous three fibres.
NDI$_S$ and BTT$_F$ SOAP spectra are nearly superimposed, with a $d_{\mathrm{SOAP}}$ $\sim0$, and very similar PCA projections (Fig.\ref{fig:fig3}A-B). This indicates that the monomeric environments associated to these systems are, on average, structurally/dynamically very similar (see also population histograms and interconversion graphs in Fig. \ref{fig:fig2}).
Conversely, BTT$_{5F}$ and BTA$_{W}$ present unique features, distancing themselves from the other fibres. Such a peculiarity, already suggested by the fingerprints of Figs.\ref{fig:fig1} and \ref{fig:fig2}, is here confirmed quantitatively. 

Classifying different types of supramolecular polymers, these results demonstrate how the proposed approach is effective in comparing 1D assemblies in general. This is possible through the unbiased, data-driven SOAP-based metric, that quantitatively compares the average molecular environment (and defects) emerging in the various assemblies. 
The flexibility of the approach suggests to investigate whether such a "defectometer" can be used to compare assemblies with higher structural dimensionality (\textit{i.e.}, 2D or 3D assemblies).

\subsection{Comparing 2D dynamic assemblies}

We here extend our approach to compare 2D supramolecular systems. We focus on assemblies where the monomers may be arranged on flat, as well as on curved surfaces. As case studies we chose DPPC lipids, which self-assemble into 2D (planar) lipid bilayers, and DPC and SDS surfactants, that form nearly spherical micellar aggregates. For all these systems, we employed validated Martini-based CG models (same resolution of the previously studied models) and collected equilibrium MD trajectories for the SOAP analysis (Fig. \ref{fig:fig5}A).
Also in these analyses, we consider one SOAP center per-monomer (centered in the lipid/surfactant heads). 

\begin{figure}[h!]
    \centering
    \includegraphics[width=0.9\textwidth]{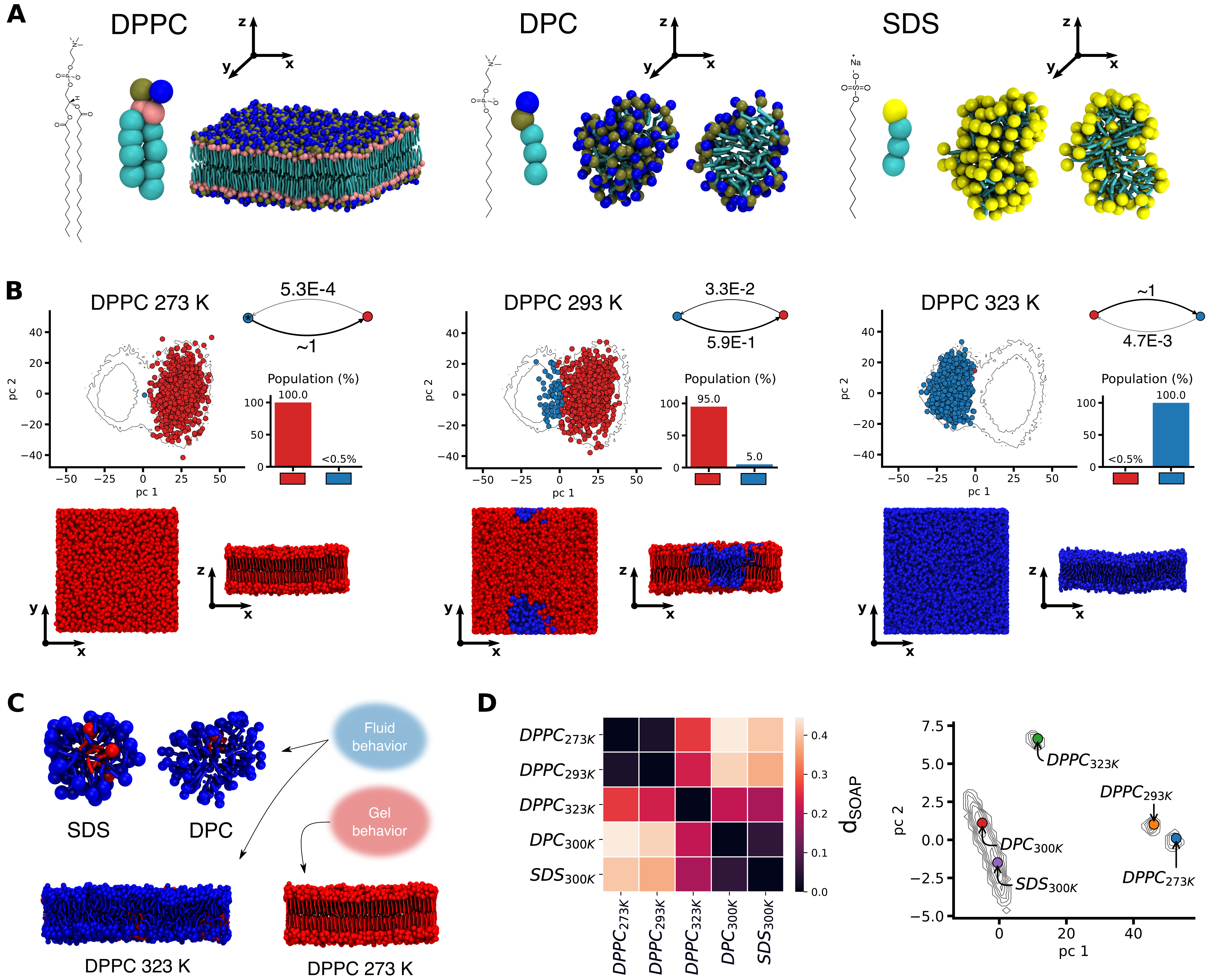}
    \caption{Comparing 2D assemblies. We compare assemblies where the monomers are displayed on flat (\textit{i.e.}, DPPC lipid bilayers) or curved/spherical surfaces ((\textit{i.e.}, DPC and SDS micelles). (A) Three example 2D assemblies (CG models for the monomers and their respective aggregates) considered herein. (B) SOAP+PAMM analysis for a DPPC lipid bilayer model at three different temperatures. For each value of $T$ we report: the PC scatter plot of SOAP feature vectors (top-left), the molecular motifs (macro-clusters) interconversion graph and population histogram (top-right), an equilibrium MD snapshot of the bilayer from top and side views (bottom). The scatter plot circles and the lipids in the snapshots are coloured according to the different motifs detected by PAMM analysis. (C) Equilibrium MD snapshots for each studied systems, with monomers coloured according to the molecular motifs (macroclusters) detected in the analysis of the global dataset of the 2D aggregates (see Fig. S9 %\ref{fig:figSI2Dassemblies}
 for the individual PCA scatter plots). (D, left) Distance $d_{\mathrm{SOAP}}$ matrix for all 2D assemblies. (D, right) Contour plot of the \textit{frame}-average distributions, computed from the PCs of the SOAP vectors. The coloured circles represent the \textit{simulation}-averages for all considered assemblies. 
    }
    \label{fig:fig4}
\end{figure}

DPPC bilayers are known to undergo a gel-to-liquid transition around $\sim300-320$ K of temperature, which is well-captured by DPPC Martini models.\cite{capelli2021} In our analysis, we considered equilibrium MD trajectories for DPPC at $T=273$ K, $T=293$ K and $T=323$ K, as representative of 2D planar assemblies having variable reconfiguration dynamics of monomers, depending on the bilayer gel/liquid state. 
Analysing the DPPC trajectories at different temperatures, we ran a first SOAP+PAMM analysis, analogous to those for comparing fiber variants that belong to the same family, in Figs. \ref{fig:fig1} and \ref{fig:fig2} (see Methods for details). The results prove that this approach captures the bilayer gel-to-liquid transition correctly, solely based on how the environment surrounding the lipids changes with the temperature (Fig. \ref{fig:fig4}B). At low temperature ($\sim273$ K) the bilayer is entirely in the gel phase (in red), while as the transition temperature is approached (between $\sim300-320$ K), liquid domains appear (in blue, $\sim5\%$ at $\sim293$ K).\cite{capelli2021} When the temperature raises to $\sim323$ K, the bilayer is entirely liquid, with residual ($<5\,\%$), gel-like lipids (Fig. \ref{fig:fig4}B: right). Our analysis distinguishes gel and liquid domains in planar lipid bilayers, solely based on the SOAP data, and without prior knowledge on the lipid arrangement in each phase. Nonetheless, interesting questions arise on how similar are the identified monomeric environments to, \textit{e.g.}, those present on the curved surface of a micelle.

 %these results follow closely ones that were obtained recently in a paper from our group.\cite{capelli2021}\cla{Sono gli stessi del paper?} \gar{No perché sono fatti in modo diverso (intendo il trattamento di SOAP e il numero di membrane), ma in modo qualitativo troviamo lo stesso trend}

We thus extended our SOAP+PAMM analysis to curved 2D assemblies, enriching the dataset with the SOAP data obtained from the equilibrium MD (at $T=300$ K) of two micellar aggregates, made of DPC or SDS surfactants. %Also in such analyses we consider one SOAP per-monomer, centered in these cases on the lipid/surfactant heads. 
The PCA scatter-plots (Fig.~S9) %\ref{fig:figSI2Dassemblies})
 indicate a significant overlap of both micelles fingerprints with that of the lipid membrane in the liquid phase ($T=323$ K).
This suggests that the structural/dynamical features of the monomeric environments characterizing these micelles are closer to those of a dynamic/liquid, rather than of a static/gel-like flat bilayer.

To obtain quantitative insights we then computed the $d_{\mathrm{SOAP}}$ matrix (Fig.~\ref{fig:fig4}D, left), and projected the SOAP \textit{frame}- and \textit{simulation}-averages of these systems along the first two PCs (Fig.~\ref{fig:fig4}D, right). The DPPC$_{273K}$ and DPPC$_{293K}$ assemblies appear relatively close to each other, and separated from the other three systems --- while in DPPC$_{293K}$ both gel and liquid phases coexist, most of the lipids are gel-like.
Similarly, DPC$_{300K}$ and SDS$_{300K}$ micelles have $d_{\mathrm{SOAP}}\sim0$, appearing close in the scatterplot (Fig. \ref{fig:fig4}D, right).
Interestingly, the liquid DPPC$_{323K}$ shows ``intermediate" features, being closer (in terms of monomers environment, disorder/reshuffling) to dynamic micelles formed by different monomers (SDS/DPC), than to the DPPC bilayer at lower temperature (Fig.\ref{fig:fig4}D).

\subsection{Comparing 3D dynamic assemblies}

\begin{figure}[h!]
    \centering
    \includegraphics[width=0.9\textwidth]{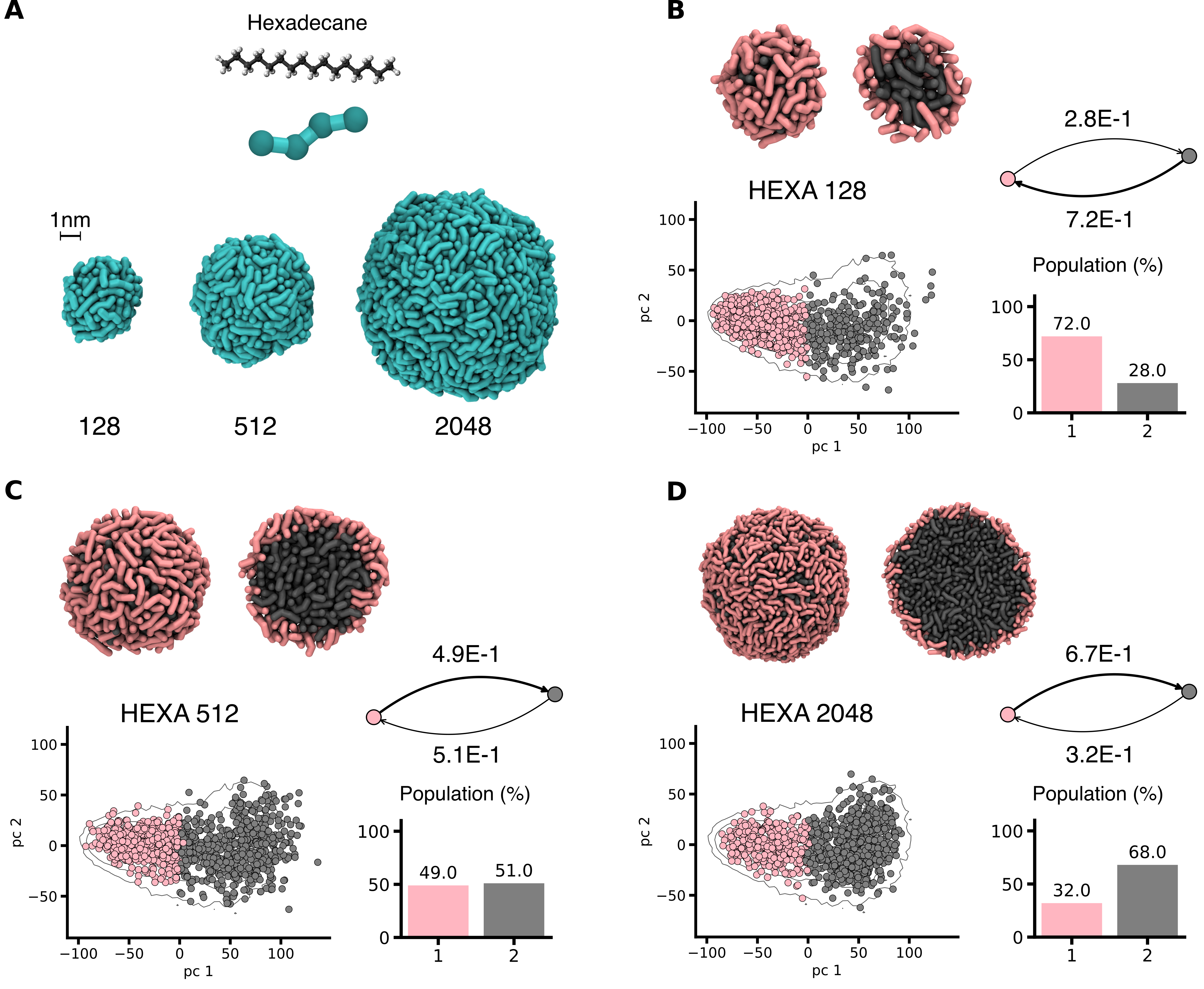}
    \caption{Comparing 3D assemblies. We consider three homogeneous nanoparticle-like structures of different size (composed of a different number of HEXA monomers). (A) Chemical structure and CG model representation of HEXA monomers and growing size nanoparticles. (B-D) SOAP+PAMM analysis of the three spherical assemblies formed by a different number of monomers (B: 128, C: 512 and D: 2048 HEXA). Each panel shows (left to right) an equilibrium MD snapshot of the aggregate (entire and in section), the PCA scatter plot of SOAP feature vectors, the molecular motif interconversion graph and the macroclusters population histogram. The colours refer to the molecular environment states detected by PAMM.} 
    \label{fig:fig5}
\end{figure}

We tested this approach also on soft self-assembled 3D assemblies. In particular, we focused on 2D nanoparticles.
As an example, we chose hexadecane (HEXA), an hydrophobic alkane composed of 16 Carbon atoms, for which we employed a Martini-based CG model. HEXA molecules undergo aggregation forming spherical assemblies (droplets) in water. We tested our method by comparing assemblies of variable size, composed of 128, 512 and 2048 molecules (Fig. \ref{fig:fig5}A). For each assembly, we collected equilibrium MD trajectories in explicit water, and analysed them \textit{via} SOAP+PAMM analysis. We again defined one SOAP vector in the center of each monomer. Since the systems compared have different size, we adapted the sampling statistics, considering different number of frames for each aggregate, according to its size (with fixed sampling frequency of $1\,\mathrm{ns}^{-1}$). Therefore, all systems contribute to the global SOAP dataset in equal percentages (see Methods for details).

The comparison of HEXA droplets is reported in Figs. \ref{fig:fig5}B-D. 
Not surprisingly, the molecular environments do not change radically across different system sizes (scatter plots in Figs. \ref{fig:fig5}B-D). Our analysis identifies two populated motifs, essentially corresponding to the monomers that are on the surface (%dynamic, in 
pink) or in the bulk (%static,
gray) of the HEXA assemblies (in general, the larger is the assembly, the more populated is the bulk phase).
The analysis also highlights a dynamicity between these two molecular motifs (interconversion between bulk and surface), which varies with the aggregate size in the same way as the ratio between the population of the pink and gray domains. 
While these cases are rather simple, they show that our analysis provides robust and reasonable results also in the case of 3D assemblies of variable size. These also enrich our database with isotropic assemblies, providing additional data to push the limits of the comparability in the next section.

\subsection{A "defectometer" to compare and classify different types of soft, dynamic assemblies}

\begin{figure}[h!]
    \centering
    \includegraphics[width=0.9\textwidth]{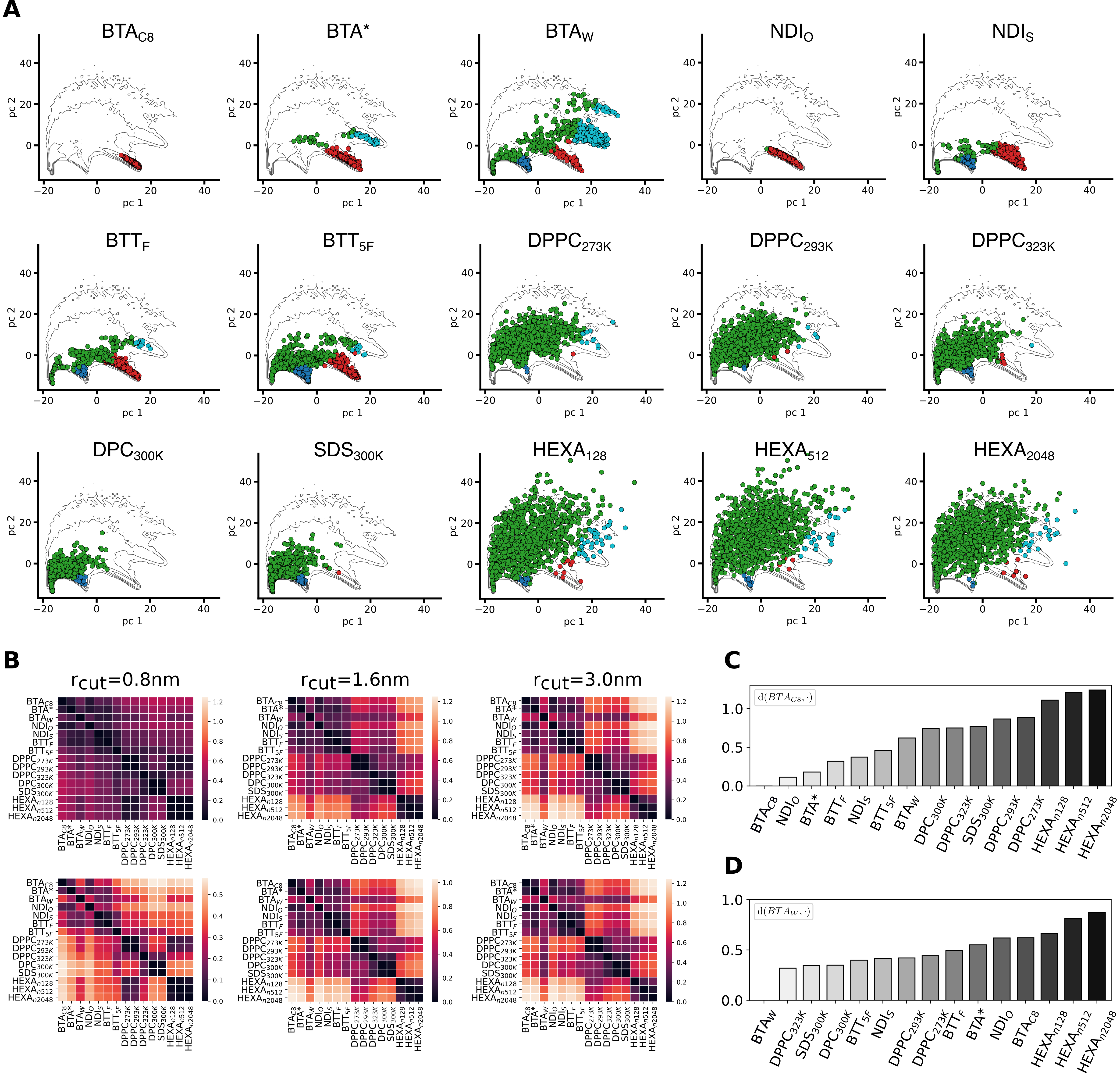}
    \caption{Comparison different types of soft dynamic supramolecular materials. (A) SOAP+PAMM analysis for all the 15 systems considered in the present work. Each panel reports the PCA scatter plot of SOAP feature vectors relative to a single system projected onto the global SOAP vector dataset (accounting for all considered systems: black contour plot). The colours indicate the molecular motifs detected by the PAMM clustering algorithm (the data obtained using a \texttt{rcut}$=0.8$ nm in the SOAP analysis are reported). (B) Distance $d_{\mathrm{SOAP}}$ matrices, computed from the SOAP \textit{simulation}-averages using different cutoffs in teh SOAP analysis (left: \texttt{rcut}$=0.8$nm, center: \texttt{rcut}$=1.6$nm, right: \texttt{rcut}$=3.0$nm); each matrix is reported twice, using a fixed $d_{\mathrm{SOAP}}$ scale (top: $0.0<d_{\mathrm{SOAP}}< 0.22$), or a $d_{\mathrm{SOAP}}$ scale that is adapted for each case (showing consistent similarity results at all \texttt{rcut} values). (C) Plot of $d_{\mathrm{SOAP}}$ between BTA$_{C8}$ (reference) and all the other assemblies (reported data computed with \texttt{rcut}$=1.6$nm). (D) Same as (C) but setting BTA$_{W}$ as the reference.} 
    \label{fig:fig6}
\end{figure}

As a last step, we processed the simulation data of all the studied systems in a single SOAP+PAMM characterization, assessing to what extent completely different assemblies can be compared to each other.
It is worth noting that all self-assembled systems share a common feature, in that they are formed of individual monomers that interact with each other. Having used a single-center per monomer SOAP definition in each system, we can thus compare all these different assemblies in a common SOAP feature space containing information on the mutual displacement of their monomers along the equilibrium MD trajectories.  %Since in our analysis we consider one SOAP center per-monomer, all different considered assemblies are thus comparable in terms of the spatial/temporal displacement of the monomers in their structures. 
We gathered in a single dataset all SOAP vectors computed in the analyses of the previous sections. An identical number of SOAP vectors is considered for each system,  to equally weight them to guarantee a balanced comparability.

We processed the complete SOAP dataset \textit{via} the described PCA+PAMM workflow. Fig.~\ref{fig:fig6}A shows the PCA projections of the SOAP vectors for each system, onto the contour map of the global dataset. Different colours are associated to the five dominant molecular motifs detected \textit{via} PAMM. %at fixed cutoff of $0.8$nm.
Different systems populate different regions in Fig.~\ref{fig:fig6}A. This comparative analysis provides results compatible with those of previous analyses (\textit{e.g.}, different PCA scatter-plots between ordered and disordered fibres). Surprisingly, many systems populate all the identified motifs, suggesting that structural analogies are present, despite the intrinsic diversity of the considered assemblies.
Nonetheless, the cluster superposition in the PCA scatter-plots hampers the comparison, as relevant dimensions might be hidden. %makes it difficult to compare the systems correctly, and this may even lead to erroneous considerations.

To obtain a more quantitative insight, we turned again to the average SOAP vectors associated to each system. Using the high-dimensional SOAP metric\cite{capelli2021} introduced above, we computed the $d_{\mathrm{SOAP}}$ matrix comparing all assemblies with each other, indicating the similarity and differences between the systems in terms of monomer environments (distance in the SOAP space). %As in the previous sections, dark colours in the off-diagonal entries of the $d_{\mathrm{SOAP}}$ matrix of Fig.~\ref{fig:fig6}B indicate reduced $d_{\mathrm{SOAP}}$ values and high similarity between two assemblies (in terms of monomer environments). Light colours identify systems which are, on the contrary, different in this sense, and which are separated by larger $d_{\mathrm{SOAP}}$ in the SOAP feature space. 
An important parameter for the SOAP analysis is the cut-off radius ($\texttt{rcut}$), which determines the size of the neighborhood considered in characterizing the molecular environment of each SOAP center.\cite{gasparotto2019identifying}  
In principle, being the SOAP metric fully data-driven, the $d_{\mathrm{SOAP}}$ scale may change while changing the $\texttt{rcut}$ in the analysis. 
In order to prove the robustness of our analysis, and given the differences in the assemblies compared herein, we computed the $d_{\mathrm{SOAP}}$ matrix using three different $\texttt{rcut}$ values: $0.8$ nm, $1.6$ nm, and $3.0$ nm. Shown in Fig.~\ref{fig:fig6}B (top panels), the matrices show a reduced $d_{\mathrm{SOAP}}$ between the various systems when a shorter $\texttt{rcut}$ is used (\textit{i.e.}, the differences between systems fade for $\texttt{rcut}=0.8$ nm with respect to $\texttt{rcut}=3.0$ nm). 
The bottom panels of Fig.~\ref{fig:fig6}B report the same $d_{\mathrm{SOAP}}$ matrices of the top panels, with an adapted colour scale. Aside from subtle differences --- \textit{e.g.}, a slightly lower resolution when a smaller $\texttt{rcut}$ is used ---, the global picture in terms of similarity between the compared assemblies is consistent in all cases (more details on the effect of changing SOAP+PAMM parameters on the resolution and completeness of the analysis are reported both in the Methods and SI).

In general, in the $d_{\mathrm{SOAP}}$ matrices we observe four main dark areas - \textit{i.e.}, that of fibers (1D assemblies), two neighboring/entangled ones for the flat and spherical 2D assemblies, and a third one %, quite separated from the other two, 
for the 3D aggregates (bottom-right in the matrix). An interesting exception is the BTA$_W$ fibre, found more similar to 2D assemblies (DPPC and surfactant micelles) than to the other ordered 1D assemblies. Another interesting system is the DPPC bilayer at $323$ K, which is found closer to highly dynamic SDS and DPC micelles rather than to DPPC at lower temperature ($293$ or $273$ K).

In the $d_{\mathrm{SOAP}}$ matrix, one can select one assembly and rank all the others with respect to it. For example, selecting the ordered BTA$_{C8}$ fibre, the plot of Fig. \ref{fig:fig6}C shows a high similarity (small $d_{\mathrm{SOAP}}$) with the other 1D ordered assemblies, lower similarity with disordered 1D fibres (\textit{e.g.}, $d_{\mathrm{SOAP}}\sim0.6-0.7$ vs. BTA$_W$), while $d_{\mathrm{SOAP}}$ increases further with respect to 2D and 3D assemblies. As anticipated above, an interesting result is obtained for BTA$_W$ (Fig.~\ref{fig:fig6}D). In terms of $d_{\mathrm{SOAP}}$ ranking, the closer assemblies to this water-soluble, disordered fibre are indeed highly dynamic,  planar or spherical 2D assemblies. Surprisingly, all the ordered 1D fibres are less similar to BTA$_W$ than all the 2D studied systems. This suggests that the solvophobic component of the BTA$_W$-BTA$_W$ interactions in water (key factor in controlling the defect formation in such 1D assemblies)\cite{gasparotto2019identifying,demarco2021controlling} can
shape a molecular environment in the surrounding of the monomers that is closer to the environment of 2D micelles or liquid-like lipid bilayers than to the environment of ordered BTA variants (\textit{e.g.}, BTA$_{C8}$). Noteworthy, this is known to produce a dynamic surface adaptability in this specific fibre that is similar to the surface fluidity seen, \textit{e.g.}, in lipid bilayers.\cite{bochicchio2017natcommBTA,torchi2018dynamics}

It is worth noting that such similarity measurements rely solely on data extracted from equilibrium MD simulations, with no major assumption on the structure/features of the studied materials. %or on the parameters used for the analysis. %to classify them according to \textit{a priori} assumptions. 
While this approach is flexible, and can be used to compare in an unbiased way different types of materials, it also creates the important opportunity to classify assemblies based on the molecular environments that populate them, which is a crucial step towards the rational design of supramolecular materials with programmable dynamic properties.

\section*{Conclusion}
Rigorously and precisely classifying the structure and dynamics of soft supramolecular materials is a crucial step towards the rational design of monomers that can self-assemble into supramolecular structures with controllable complex dynamic behaviors. While it is known that the dynamics of such materials originate from defects that may be present or may form in their assembled structure, designing a unified, unbiased, and robust approach to classify soft assemblies based on their "defectivity" is not trivial.   

Here we report an unsupervised machine learning approach that allows comparing and classifying soft self-assembled materials (1D supramolecular fibres, 2D, and 3D assemblies) based on the structural dynamics of the internal molecular environments that surround their monomers. We analyse equilibrium MD trajectories of the assemblies \textit{via} a synergistic use of atomic environment descriptors, unsupervised clustering, and similarity/dissimilarity measurements between the compared systems in a high-dimensional feature space. We use molecular models for the assemblies having the same submolecular ($<5$ \AA) resolution in the representation of the systems. This guarantees sufficient resolution in our analysis and comparability of the results. Monomers' displacement/arrangement data extracted from the MD simulations are translated into relevant information on the molecular arrangements within the assemblies by means of SOAP feature vectors.\cite{bartok2013} The high-dimensional data contained in the resulting SOAP spectra are then used to identify dominant monomeric clusters/states in the various assemblies \textit{via} unsupervised PAMM clustering (and PCA).\cite{gasparotto2018PAMM,gasparotto2019identifying} The analysis outputs a characteristic fingerprint for each system in terms of the most populated molecular motifs, their structural and dynamical features. A metric defined in the high-dimensional SOAP feature space\cite{capelli2021,capelli2021arx} then provides a quantitative way to compare and classify of the studied systems. 
Here we show how such data-driven analysis can be used to compare a wide range of supramolecular systems, providing a classification that is entirely based on data on the monomeric motifs that emerge within them at the equilibrium. The obtained results also demonstrate how, monitoring the monomer transitions between the detected molecular motifs, it is possible to obtain relevant information on the inner dynamics of the assemblies.
We can observe how ordered 1D stacked fibres are quite similar with each other, while defected supramolecular polymers (\textit{e.g.}, BTA$_W$) have a richer and diverse internal structure/dynamics, closer to that of some types of considered lipid bilayers or micelles. This fits well, \textit{e.g.}, with the complex dynamics of the surface of BTA$_W$ water-soluble fibres,\cite{albertazzi2014,baker2016,lou2017,bochicchio2017natcommBTA} and with their dynamic adaptivity and structural reconfigurability.\cite{albertazzi2013,torchi2018dynamics} 

This method is powerful for multiple reasons. (i) It does not build on any \textit{a priori} knowledge on the structure and dynamics of the various assemblies that are compared. (ii) Building on the concept of "defectivity", it proposes defects as a common ground to compare between materials, which holds a great potential to unify supramolecular materials. (iii) Such a data-driven "defectometer" allows to quantitatively classify dynamic assemblies that are different from each other (\textit{e.g.}, fibres \textit{vs.} micelles \textit{vs.} layers \textit{vs.} nanoparticles). This provides us with a precious tool towards the rational design of self-assembled materials with controllable dynamic properties, which is key to conceive complex systems where multiple assembled entities can effectively communicate with each other in a dynamic way.

\section*{Methods}\label{sec.methods}

\subsection*{Descriptors of atomic environments}

Let a generic output of an MD simulation be the ensemble of system conformations sampled at a series of MD timesteps, represented by the atomic coordinates $\mathbf{R}(t)$ of the molecular species in the system at timestep $t$. $\mathbf{R}$ is a $3N$-dimensional configuration vector, where $N$ is the number of particles. %$A = \{\mathbf{r}_i, \alpha_i\}$, with $\mathbf{r}_i \in \mathbb{R}^3$, the three dimensional space.
A generic descriptor is a mapping from the $3N$-dimensional coordinate space to a $D$-dimensional \textit{feature space}, that associates a \textit{feature vector} to each of the sampled conformations $\mathbf{R}(t)$. %
We aim at a descriptor capable to characterize and compare the atomic/molecular environment surrounding the multiple constituents of a soft, supramolecular system.
A first, crucial requirement of this mapping is that it preserves physical symmetries such as permutation, translation and rotation invariance, ensuring that physically equivalent configurations are recognized as such by the descriptor.\cite{ceriotti2019understading}
Other less obvious properties, required for a useful representation of the atomic/molecular environment are \textit{e.g.} smoothness, additivity and level of locality.\cite{musil2021,musil2021physicsinspired}
SOAP descriptors satisfy all these requirements.

\subsubsection*{Smooth Overlap of Atomic Position (SOAP)}
The Smooth Overlap of Atomic Position (SOAP)\cite{bartok2013} is an  atom-centered descriptor that accurately reproduces many-body density correlation features of many-body systems.
SOAP were introduced in Ref.~\citenum{bartok2013} as new bond-order parameters, able to efficiently include radial and angular information of the environment that surrounds atoms or molecules.~\cite{bartok2013,musil2021,deringer2021}
The SOAP descriptor of a system of $N$ particles describes the atomic/molecular surroundings of a selected set of $M$ coordinates of the system components, which are referred to as the ``centers" of the SOAP vector.
These $M$ centers can include the position of every single atom of the system, as well as selections or combinations (as \textit{e.g.} center of geometry/mass) of them. In the following we detail the choice of the SOAP center for each of the studied systems, but in general we associate a single SOAP vector per each monomer (see also Sec.~\ref{sec:res}).
In this work we have considered only single-specie systems, %(\textit{i.e.} $\alpha_i = \alpha$), 
but the approach is generalizable to multiple species.

The SOAP descriptor is built from the density of neighboring centers $j$ that surround the $i$-th center, namely
\begin{equation}
    \label{eq:densiset}
    \rho^{i}(\mathbf{r}) = \sum_j \exp \left[ \frac{- | \mathbf{r} - \mathbf{r}_{ij} |^2 }{ 2\sigma^2 } \right] f_{\texttt{rcut}}(| \mathbf{r} - \mathbf{r}_{ij} |),
\end{equation}
where a Gaussian function is associated to each neighboring center (located at distance $\mathbf{r}=\mathbf{r}_{ij}$ from the $i$-th center), to build a smooth density function. The $\sigma$ parameter sets the width of the Gaussian located at the $j$-th neighboring center. The total neighbor density $\rho = \sum_i \rho^i$ is retrieved by summing all contributions.
The function $f_{\texttt{rcut}}$ smoothly goes from 1 to 0 at $\texttt{rcut}$, so that the environment of each center extends up to a fixed cutoff $\texttt{rcut}$. % (an input parameter of the descriptor). 
Starting from Equation \ref{eq:densiset}, which is intrinsically invariant for the permutation of centers, the SOAP descriptors are defined by incorporating the translation and rotation invariances.
This is done in two steps: (i) Eq.~\ref{eq:densiset} is expanded in the basis of orthonormal radial functions ${R_n(r)}$ and spherical harmonics ${Y_{l,m}(\hat{\mathbf{r}})}$; (ii) rotational invariance is enforced by building symmetrized combinations of the expansion coefficients.\cite{bartok2013,musil2021}

We here employ the second-order SOAP descriptor, also called SOAP power-spectrum (in analogy with the Fourier analysis), which can be written as
\begin{equation}
    \label{eq:powerspectrum}
    \gamma^{i}_{nn'l} \propto \frac{1}{\sqrt{2l+1}} \sum_{m=-l}^{+l} ( c^{i}_{nlm} )^* c^{i}_{n'lm},
\end{equation}
where $c^{i}_{nlm}$ are the expansion coefficients of the particle density surrounding the $i$th-center.
The full SOAP descriptor associated to the $i$-th center is a vector including all contributions from Eq.\ref{eq:powerspectrum},
\begin{equation}
    \label{eq:fullps}
    \mathbf{p}_i = \{ \gamma^{i}_{nn'l} \} ,
\end{equation}
where $n$ and $n'$ range from $1$ to $\texttt{nmax}$ and $l$ ranges from $1$ to $\texttt{lmax}$, setting the dimension $D$ of the vector (in cases of multiple species $D$ depends also on the number of species).\cite{bartok2013,deringer2021}
To compute Eq.\ref{eq:powerspectrum} we used the python package DScribe,\cite{dscribe} using \texttt{nmax}, \texttt{lmax}$= 8$ and three different cutoff values \texttt{rcut}$=0.8, 1.6, 3.0$ nm. The remaining parameters were set to default values of the DScribe library.

To summarize, for a given $3N$-dimensional configuration vector $\mathbf{R}$ sampled via CG-MD, we define a set of $M$ centers $\{i\}$, and for each center we compute the SOAP vector $\{ \mathbf{p}_i \}$. 
This generates a dataset of $M$ SOAP vectors that describe the structural arrangement of the centers in the selected configuration of the system. Such SOAP vectors represent ``local" descriptors, encoding the information on the environments that surrounds each center.

We can also define ``global" descriptors, useful for the comparison of different systems: 
(i) the \textit{frame}-average SOAP descriptor $\bar{\mathbf{p}}_t = \{ \bar{\gamma}_{nn'l}^t \}$, where each component of the vector is computed as the power spectrum of the density (Eq.~\ref{eq:densiset}), after averaging over all the $M$ centers, namely: \cite{dscribe}
\begin{equation}
    \label{eq:psframeave}
    \bar{\gamma}_{nn'l}^t \sim \sum_{m=-l}^{+l} \left( \frac{1}{M} \sum_i^{M} c^{i}_{nlm} \right)^* \left( \frac{1}{M} \sum_i^{M} c^{i}_{n'lm} \right) .
\end{equation}
This gives a global (averaged) picture of the structural features characterizing a specific MD frame.
(ii) The \textit{simulation}-average SOAP descriptor,
\begin{equation}
    \label{eq:pssimave}
    \langle \bar{\mathbf{p}} \rangle = \frac{1}{T} \sum_t^{T} \bar{\mathbf{p}}_t  ,
\end{equation}
that averages all the \textit{frame}-average SOAP descriptors along the $T$ frames collected through the MD simulation.
Eq.\ref{eq:pssimave} represents a compact global fingerprint for the equilibrium structure of the system under investigation, provided that the $T$ frames are sampled at the equilibrium. %takes on a physical meaning only if the structure are assumed to be sampled at equilibrium conditions, $\langle \bar{\mathbf{p}} \rangle \sim \bar{\mathbf{p}}_t$ for every $t$, so the simulation-averaged SOAP vector .
We used the \textit{frame}-average SOAP to assert similarities between different structures.

\subsubsection*{Comparison of Molecular Environments}
Once SOAP feature vectors are computed, the similarity between the SOAP characterization of different supramolecular structures can be inferred \textit{via} a distance metric (or kernel function), provided that the dimensionality $D$ of the compared SOAP vectors is the same (\textit{i.e.} using equal \texttt{nmax} and \texttt{lmax}) between the compared systems.
We measure the similarity between two SOAP vectors using a linear polynomial kernel,% , the scalar product of two vectors raised to an integer power $\xi$,
\cite{bartok2013,de2016comparing}
\begin{equation}
    \label{eq:linkernel}
    \mathcal{K}(i,j) = \left( \mathbf{q}_i \cdot \mathbf{q}_j \right), %^\xi ,
\end{equation}
where $\mathbf{q} = \mathbf{p} / |\mathbf{p}|$ is the unit-normalized SOAP vector. %We here used the so-called linear kernel ($\xi = 1$).
Upon normalization, $\mathcal{K}(i,j)$ is equal to 1 if the two molecular environments are exactly superimposed or 0 if no overlap occurs.
Since Eq.\ref{eq:linkernel} defines a positive-definite kernel, it naturally induces a metric, for which the distance between two feature vectors is defined as\cite{de2016comparing, musil2021physicsinspired}
\begin{equation}
    \label{eq:kerdistance}
    \mathrm{d}_{\mathrm{SOAP}}(i,j) = \sqrt{\mathcal{K}(i,i) + \mathcal{K}(j,j) - 2\mathcal{K}(i,j)} .
\end{equation}
Recently, this SOAP-induced metric was employed in Ref.~\citenum{capelli2021} to compare and classify the structural arrangement of lipid membranes represented \textit{via} different force-fields.
We here apply the same approach for the comparison of different supramolecular systems, computing Eq.~\ref{eq:kerdistance} for the pairs of \textit{simulation}-average SOAP vectors obtained \textit{via} the simulation of each different test case.

The SOAP metric of Eq.~\ref{eq:kerdistance} fully depends on the high-dimensional information contained in the SOAP spectra, and it provides a classification that is free from prior assumptions on the structural order of the considered systems. 
Nonetheless, the setting of the cutoff radius \texttt{rcut} in the calculation of the SOAP can influence the resulting feature vectors, and few considerations are useful in this sense. The comparative analysis of all the systems, reported in Fig.\ref{fig:fig6}, shows that the choice of a smaller cutoff ($\texttt{rcut}=0.8$ nm) can lower the accuracy of the comparison, especially when substantially different systems are compared (see the $d_{\mathrm{SOAP}}$ matrices in Fig.\ref{fig:fig6}B:top, and the SI for additional discussion on this matter).
Nonetheless, even a $\texttt{rcut}=0.8$ nm (which takes into account only the closest neighbors of the monomers in the SOAP analysis) appears sufficient to obtain a good characterization of the molecular environment in each system, capturing an analogous pattern of relative distances $d_{\mathrm{SOAP}}$ with respect to the results obtained with larger $\texttt{rcut}$ (see the rescaled  $d_{\mathrm{SOAP}}$ matrices in Fig.\ref{fig:fig6}B:bottom).
We underline that we chose a fixed value of $\texttt{rcut}$ for all the systems, in each of the analysis reported previously.
This is made possible by the fact that the studied systems are all described with a similar resolution, namely employing Martini-based CG models.
This gives rise to comparable inter-beads distances and facilitates a comprehensive analysis that accounts for the same neighborhood area per each monomer.
In principle it should be possible to include in the dataset also simulation results obtained with molecular models at different resolution, provided that the cutoff is differentiated to have a consistent SOAP description across the different models. 

\subsection*{Dimensionality reduction}
The SOAP data output of an MD simulation consists of a set of feature vectors (Eq.\ref{eq:fullps}) of dimension $D$ (typically large).
High-dimensional data, while rich in information, are not well-suited for visualization and classification/clustering, especially with methods based on Euclidean distance. %, and  to visualize.
To overcome these limitations a dimensionality reduction method is necessary.
In this work we employed (linear) Principal Component Analysis (PCA)\cite{Pearson1901PCA,Hotelling1933PCA} both as a way to pre-process the SOAP data, prior to the unsupervised clustering step, and for simple visualization purposes.
PCA projects the ensemble of SOAP data %matrix $\mathbf{G}$ in terms of a \textit{latent space} matrix $\mathbf{T} = [\ N_{\mathrm{vec}} \times N_{\mathrm{PCA}} ]\ $, where $N_{\mathrm{PCA}} \ll D$.
onto an orthogonal basis set, that best accounts for the variance in the original dataset, namely catching the main directions along which the SOAP feature vectors undergo the larger variations. The first components of this projection contain most of the relevant information, allowing one to retain only few components, thus reducing the dimensionality.
We here performed PCA \textit{via} the python class \texttt{sklearn.decomposition.PCA()} from the python library Scikit-learn\cite{scikit-learn}.
After PCA, we chose to keep only three principal components, (if not stated otherwise), as this  allows us to maintain more than 80\% of the total variance in all our cases (see Fig.~S3%\ref{fig:figSIpcvariance}
).
The PAMM clustering step is performed on 3-dimensional vectors containing the first three PCs of the SOAP feature vectors.
For visual representation of the SOAP vectors in the scatter-plots (Figs.~\ref{fig:fig1}-\ref{fig:fig6}) we employed the first two PC projections.

\subsection*{Building of a training set}
Crucial to our comparative analysis is the construction of a ``shared'' dataset, that comprises SOAP feature vectors from all the individual datasets associated to the systems that are compared.
This shared dataset is then used to train the PCA model reducing the dimensionality of the SOAP feature vectors, so that we can proceed with the application of the clustering algorithm and visualization of the data in the PC-plane.
The usage of comprehensive data coming from a set of multiple systems allows us to take into account the overall data diversity, and compare the molecular environment across the different systems.
To allow for a faster PCA step, we sampled the production trajectories including equal subsets of conformations per each system. % did not use the entire sampling obtained by the MD production runs, but randomly sampled representative data subsets.

\subsection*{Unsupervised clustering}
Relevant patterns in the SOAP data are detected by means of unsupervised clustering:
we used Probabilistic Analysis of Molecular Motifs (PAMM)\cite{gasparotto2018PAMM}, a density-based clustering algorithm, specifically developed to partition the SOAP data collected \textit{via} MD simulations.
In Ref.~\citenum{gasparotto2018PAMM} the authors presented the complete workflow for the algorithm. %, which was later successfully used to completely characterize the structural behavior of BTA-based supramolecular polymers\cite{gasparotto2019identifying, demarco2021controlling}.
The algorithm takes as input a set of $N$ feature vectors, that represent local or global SOAP descriptors. As detailed earlier, in our case the algorithm processes the three-dimensional projections of SOAP vectors on the first three PCs. Then, the Probability Distribution Function (PDF) of these dimensionally reduced SOAP vectors is estimated by means of a Kernel Density Estimation (KDE), built from a multivariate anisotropic Gaussian function in the 3D space. In order to mitigate the computational load of KDE, the PDF estimation is done on a grid of $N_{\mathrm{grid}} \subset N$ points, selected non-uniformly through a farthest point sampling method.
After the PDF is accurately estimated a density-based clustering algorithm, based on Gaussian Mixture Modeling (GMM), associates each different local maximum in the PDF to a cluster.
Once this analysis is completed we obtain the so-called Probabilistic Motifs Identifiers (PMIs) (as the molecular motifs were called in the original reference)\cite{gasparotto2018PAMM} that characterize the features of the system at hands.
All clustering analyses were conducted using the original PAMM algorithm (Available online at https://github.com/cosmo-epfl/pamm) modified with a tailored Python3 wrapper for handling the different analysis steps and post-processing.

\subsection*{Molecular Dynamics simulations}

All the simulations presented in this work were performed using the MD package GROMACS\cite{abraham2015gromacs}, version \texttt{2018.6}, from the setup of the simulation box, to the equilibration and production runs.
We adopted CG descriptions for each system, built \textit{via} the Martini scheme\cite{marrink2007martini}, with explicit solvent description. The parametrization of the monomer models was conducted following the literature data (where available), using the standard, non-polarizable Martini force-field\cite{marrink2007martini} (version 2.2). Details per each system are provided in the following.
All simulations were performed using Periodic Boundary Conditions (PBC) to limit the finite-size effects, also allowing to simulate portions of infinite aggregates or membranes.
To prepare the CG model systems for MD in equilibrium conditions we performed CG-MD equilibration runs of the order of $\mu$s in NPT conditions, starting from energy-minimized system conformations. 
We then proceeded with the production CG-MD runs, to collect the statistics of equilibrium configurations processed in our analyses; all our systems where sampled every 1 ns of CG-MD. %, aside from the DPPC lipid membranes that where sampled every 10 ns.
For all the CG-MD runs (equilibration and production) we used a 20 fs time-step, an interaction cutoff (1.1 nm) where the non-bonded potentials are truncated and shifted (consistently with the Martini scheme). The Verlet neighbor list scheme\cite{deJong2016} is employed to reduce the load of calculating pair interactions.
The temperature was maintained through the V-rescale thermostat,\cite{Bussi2007} set at 300 K for all the simulations, except the cases where a different temperature is stated (\textit{i.e.} for lipid membranes) with a coupling constant of 1.0 ps.  
The pressure was maintained \textit{via} Berendsen barostat\cite{Berendsen1984}, set at 1 atm with a coupling constant of 2 ps.
Isotropic scaling of the simulation cell is adopted when finite aggregates are simulated (\textit{e.g.} the HEXA clusters), while semiisotropic scaling (decoupling $x/y$ from $z$) is adopted when a portion of an infinite aggregate is considered (crossing the PBCs along one or two axes), as for fibers and membranes.
In the following, we report the model references and the specific simulation setup for each system studied.

\subsubsection*{Supramolecular polymers (1D-assemblies)}
The supramolecular polymers studied in this work belong to three families, characterized by a specific chemical structure of their monomer functional core. For each family we considered monomer variants:
\begin{itemize}
\item 1,3,5-Benzenetricarboxamides (BTA) monomers, already extensively studied in the literature\cite{bejagam2015bta,bochicchio2017natcommBTA,gasparotto2019identifying}.
We considered three different variants: a water soluble BTA$_W$ (Fig.\ref{fig:fig1}A), an organic solvent (\textit{e.g.}~octane) soluble BTA$_{C8}$ (Fig.\ref{fig:fig1}B), and an intermediate case, BTA$^*$ (Fig.\ref{fig:fig1}C), obtained from BTA$_{C8}$ by artificially changing the inter-monomer interaction.
The parametrization used for the three monomer models was the same as in Ref.~\citenum{gasparotto2019identifying} and the solvents, water and octane, were parametrized according to Martini standards\cite{marrink2007martini}.
For the calculation of SOAP vectors we associated a single center for each BTA monomer, namely the center-of-geometry (COG) of the monomer core (the three central beads in Fig.~\ref{fig:fig1}), which was proven sufficient to describe the structural dynamics of such supramolecular fibre.\cite{gasparotto2019identifying} 
\item Core-substituted naphthalene diimide (cNDI)\cite{sarkar2020NDImonomer} monomers (Fig.\ref{fig:fig2}A); we studied two different variants of cNDIs, by changing the substituent atom on the side of the core structure.
Following Ref.\cite{sarkar2020NDImonomer} we considered a substituted Oxigen (NDI$_O$, Fig.\ref{fig:fig2}A) and Sulfur (NDI$_S$, Fig.\ref{fig:fig2}B).
The parametrization of the two monomer models was that of Ref.~\citenum{sarkar2020NDImonomer}.
In both cases the solvent was cyclohexane, parametrized according to Martini standards\cite{marrink2007martini}.
Also for the analysis of NDI dynamics we selected the COG of the monomer cores (which correspond to the central pink bead in the central arrangement of five pink beads in Fig. \ref{fig:fig2}A and B) as SOAP center. 
\item Benzotrithiophene (BTT)\cite{casellas2018btt} monomers; we studied two variants of these planar, water-soluble monomers, changing the three amino-acids attached to the aromatic core of the molecule. As in Ref.~\cite{casellas2018btt} we used L-phenylalanine for the first variant (BTT$_F$, Fig.~\ref{fig:fig2}C) , and pentafluoro-L-phenylalanine for the other (BTT$_{5F}$, Fig.~\ref{fig:fig2}D).
Both monomers present octaethylene glycol side-chains that impart water-solubility to the compound.
The CG parametrization of these monomers was the same of Ref.~\citenum{casellas2018btt}. The solvent used for these BTT polymers was water, parametrized according to Martini standards (Martini water\cite{marrink2007martini}).
The COG of the monomer cores (computed amongst the 3 central pink beads in Fig. \ref{fig:fig2}C and D) was employed as SOAP center. 
\end{itemize}

For all the seven fibre-systems we have built an ordered, one-dimensional stack that crosses the PBCs along the principal axis of the aggregate, so that the monomers at the extremes are in contact with each other mimicking and a infinite-fibre portion. Each pre-stacked aggregate was then equilibrated, and a CG-MD production run of about $\sim2\, \mathrm{\mu s}$ was performed to sample the structural dynamics in equilibrium conditions. %, of which we considered the last 2 $\mu$s for analysis purposes.

\subsubsection*{Micelles and membranes (2D-assemblies)}
The lipid molecule selected as reference case for two-dimensional aggregates was the dipalmitoylphosphatidylcholine phospholipid (DPPC, Fig.~\ref{fig:fig4}A) for which homogeneous membranes at three different temperatures ($273$K, $293$K, and $323$K, Fig.~\ref{fig:fig4}B) were prepared.
In this work we utilized the CG Martini parametrization adopted in Ref.~\citenum{capelli2021}, where SOAP+PAMM characterization of this phospholipid membrane, modelled using different descriptions, is performed.
As test cases for the micelle aggregates we selected two surfactant molecules, namely Dodecylphosphocholine (DPC), and Sodium-dodecylsulfate (SDS); both parametrized according to the standard Martini force-field.\cite{marrink2007martini,alessandri2021martini} The explicit solvent used for these lipid and surfactant systems was Martini water.
The equilibrated structure of the surfactant micelles was obtained \textit{via} spontaneous self-assembly from dispersed monomers, to obtain a size near the normal size distribution at our conditions.
The adopted equilibration procedure for the bilayer models is described in Ref.\citenum{capelli2021} (which is the minimization and equilibration protocol given by CHARMM-GUI\cite{Lee2015}).
For each system we performed production CG-MD runs of $1\ \mu$s of simulation time. 
For DPPC, DPC and SDS molecules we employed a single-center approach for the SOAP vector calculation, choosing the CG-bead that represents the head of the amphiphilic molecules as SOAP center (\textit{i.e.} the ``PO4'' (DPPC), ``PO4'' (DPC) and ``SO3'' (SDS) beads in Fig.~\ref{fig:fig4}A).

\subsubsection*{Spherical nanoparticles (3D-assemblies)}
The molecule chosen as representative case for spherical aggregates is the alkane hydrocarbon Hexadecane (HEXA); we parametrized this molecule following the Martini force-field standards.\cite{marrink2007martini} The explicit solvent used for these simulations was Martini water.
Three aggregate structures of different size (128, 512 and 2048 monomers) were constructed \text{via} 2 $\mu$s of CG-MD equilibration run, starting from randomly dispersed monomers in water. During such CG-MD stage the monomers quickly self-assemble and equilibrate forming the spherical structures represented in Fig.\ref{fig:fig5}A). 
Production runs of different length were carried out to gather the data for the analysis of the equilibrium structures, collecting the same number of SOAP vectors per each systems, independently of their different sizes. Namely, the 128 monomer system was simulated for 2 $\mu$s, the 512 monomer system for 0.5 $\mu$s and the 2048 monomer system for 0.125 $\mu$s.  As center for the SOAP representation we chose the COG of the two central beads in the CG model of HEXA.

\clearpage
\bibliography{biblio}

\section*{Data availability}

Details on the molecular models and MD simulations, and additional MD data are provided in the Methods section and in the Supplementary Information. Computational materials and data pertaining to the study conducted herein are available at: https://github.com/GMPavanLab/MLdefectometer. Other information needed is available from the corresponding author upon request.

\section*{Acknowledgements}
G.M.P. acknowledges the funding received by the ERC under the European Union’s Horizon 2020 research and innovation program (grant agreement no. 818776 - DYNAPOL) and by the Swiss National Science Foundation (SNSF grants IZLIZ2\_183336). The authors acknowledge the computational resources provided by the Swiss National Supercomputing Center (CSCS), CINECA and HPC@POLITO (http://www.hpc.polito.it). 

\section*{Author contributions}

G.M.P. conceived this research and supervised the work. A.G performed the simulations and the analyses. A.G. and G.D. implemented the analysis methods. A.G. C.P., G.D. and G.M.P. discussed the results. A.G., C.P. and G.M.P wrote the manuscript.

\section*{Competing interests statement} 

The authors declare no competing interests.

\label{mylastpage}
\end{document}